\documentclass[final]{aipproc}
\usepackage{graphicx,natbib}
\layoutstyle{6x9}
\begin{document}

\title{Magnetic fields and radiative feedback in the star formation process}


\author{Daniel J. Price}{
  address={Centre for Stellar and Planetary Astrophysics, School of Mathematical Sciences, Monash University, Vic 3800, Australia}
,altaddress={School of Physics, University of Exeter, Stocker Rd, Exeter EX4 4QL} }

\author{Matthew R. Bate}{
  address={School of Physics, University of Exeter, Stocker Rd, Exeter EX4 4QL}
}


\begin{abstract}
 Star formation is a complex process involving the interplay of many physical effects, including gravity, turbulent gas dynamics, magnetic fields and radiation. Our understanding of the process has improved substantially in recent years, primarily as a result of our increased ability to incorporate the relevant physics in numerical calculations of the star formation process. In this contribution we present an overview of our recent studies of star cluster formation in turbulent, magnetised clouds using self-gravitating radiation-magnetohydrodynamics calculations\citep{pb08,pb09}. Our incorporation of magnetic fields and radiative transfer into the Smoothed Particle Hydrodynamics method are discussed. We highlight how magnetic fields and radiative heating of the gas around newborn stars can solve several of the key puzzles in star formation, including an explanation for why star formation is such a slow and inefficient process. However, the presence of magnetic fields at observed strengths in collapsing protostellar cores also leads to problems on smaller scales, including a difficulty in forming protostellar discs and binary stars \citep{pb07,ht08}, which suggests that our understanding of the role of magnetic fields in star formation is not yet complete.
\end{abstract}

\maketitle

\section{Introduction}
 Star formation is a ubiquitous phenomenon in spiral galaxies such as our very own Milky Way. However the detailed process by which gas is converted into stars in such galaxies is still relatively poorly understood. One of the key open questions is why star formation is so remarkably inefficient, with only a few percent of the mass of gas in a molecular cloud ending up in stars. Recent observational results for nearby molecular clouds in the Spitzer cores-to-discs survey \citep{evansetal09} find the mass of stars in a star-forming clouds typically around 3-6\% of the mass of the parental molecular cloud, the latter estimated by multiplying the column density inferred from interstellar dust extinction maps by the area of the cloud (defined by an extinction threshold). Equivalently the observational result can be restated as saying that only a few percent of molecular cloud gas is converted into stars per gravitational free-fall time.
 
  Thus star formation appears to be both \emph{inefficient} --- in the sense that not much gas has been converted into stars --- and \emph{slow}, in the sense that, over the timescales necessary for gravity to act on the global cloud, not many stars have formed. It is also important to note that the inefficiencies found by \citet{evansetal09} refer to nearby molecular clouds that are actively forming stars. It is a further challenge to explain molecular clouds where relatively little star formation occurs at all, such as the Pipe Nebula (efficiency $\sim 0.06\%$, \citet{forbrichetal09}) that lies in stark contrast to the profusion of star formation occurring in the nearby $\rho$-Ophiuchus cloud.
 
  The fact that star formation appears to occur on a slower timescale than the gravitational one indicates that the answer must lie in physics beyond gravity, or at least beyond the \emph{self-gravity} of the cloud. To achieve inefficiency of star formation over time scales much greater than the dynamical time must further involve involve unbinding the cloud in some way -- for example by internal driving of turbulence by jets and outflows \citep{nl07} or invoking tidal forces from the Galactic potential \citep{bpetal09}.

  Magnetic fields are, observationally, a good candidate for explaining why clouds in otherwise similar environments can have vastly different star formation efficiencies. For example, recent optical polarisation maps of the Pipe Nebula \citep{alvesetal08} reveal a remarkable degree of uniformity in the magnetic field (as inferred from the polarisation angles), in contrast to the wide dispersion in polarisation angles (on large scales) seen in active star forming regions like Orion \citep{lietal09}. The nearby Taurus molecular cloud, recently surveyed by \citet{goldsmithetal08}, forms an intermediate case, with relatively inefficient star formation (but more efficient than the Pipe Nebula) and also a well-ordered large scale magnetic field (better ordered than Orion, though less well ordered than the Pipe), together with compelling evidence for magnetic fields strong enough to control the flow of gas in (relatively) low density outer regions \citep{heyeretal08}.
  
   The effect of a magnetic field on the self-gravitating collapse of gas to form stars can be quantified in terms of the ratio of mass within a volume to the magnetic flux threading the surface of that volume. At a critical value magnetic fields are able to prevent collapse entirely, unless some decoupling of the magnetic field from the gas occurs (i.e., ambipolar diffusion). Indeed, this theoretical understanding led to the so-called `standard model' of star formation as a quasi-static diffusion process mediated by magnetic fields \citep{sal87}. However for all star formation to occur in this manner requires that all molecular cloud cores are sub-critical (magnetic fields able to prevent collapse), whereas Zeeman measurements of field strengths in cores indicate that they are generally marginally supercritical (mass-to-flux ratios of ~few times critical, see \citet{crutcher99}). A further difficulty is the problem of how to maintain the observed turbulent motions in the cloud, since supersonic turbulence decays rapidly with or without magnetic fields in the absence of a continual driving mechanism \citep[e.g.,][]{osg01}.
   
    This latter consideration in particular has led many to consider so-called `rapid' or `turbulent' models of star formation, where the main controlling ingredient is turbulence rather than magnetic fields, with clouds that assemble, form stars and disperse within roughly one crossing time \citep{elmegreen00}. Indeed, simulations based on simply the interaction of turbulent gas dynamics and gravity alone \citep[e.g.][]{bbb03,bb05,bate09a} do a remarkably good job of reproducing many observed statistical properties of star formation, including the gross characteristics of the initial mass function, multiplicity as a function of mass and the frequency of low mass binary stars \citep{bate09a}. However, the star formation efficiency in these calculations is much higher than observed, since the fraction of gas that is initially bound and will remain so in the absence of feedback processes, giving a SFE of order 50\% (the typical fraction of gas that is bound at the end of the calculations). Improved statistics in more recent calculations of larger clouds \citep{bate09a} also suggest that there is also a problem in terms of an over-production of very low mass stars and brown dwarfs.
    
    Whilst observations indicate that magnetic fields are not strong enough to prevent global collapse in typical clouds, they can nevertheless play a determining role in the internal dynamics by acting as a source of pressure within the cloud (quantified by the ratio of gas-to-magnetic pressure: The plasma $\beta$) and by the magnetic braking of rotating cores. Indeed, observationally typical values for $\beta$ in molecular cloud cores are of order $0.3$ (that is, magnetic pressure dominant over gas pressure by a factor of 3) \citep{crutcher99,bourkeetal01}, similar to the value found in the cold neutral medium thought to be the precursor of molecular clouds \citep{ht04}.

    In a recent series of papers \citep{pb07,pb08,pb09} we have studied the effect of magnetic fields on the formation of stars in precisely this regime: where the magnetic field is too weak to prevent global collapse but sufficiently strong to play an important role as a source of additional pressure and in magnetic braking of rotating cores. In addition we have combined this with an improved treatment of the thermodynamics of the gas on small scales by incorporating a full treatment of radiative transfer in the flux-limited diffusion approximation. We propose that slow and inefficient star formation can be explained by the combined effects of magnetic fields and radiative feedback on the star formation process \citep{pb09}. We give an overview of our findings below.

\section{Numerical methods}
 Modelling star formation is made difficult by the tremendous range of length and time scales involved. For example, to follow the collapse of a giant molecular cloud of size $\sim$a few pc ($\sim 10^{11}$ km) and containing up to $10^{4} M_{\odot}$ of material to a star the size and mass of the Sun ($R_{\odot}\sim 10^{5}$ km) requires resolution over 6 orders of magnitude in length, around 14 orders of magnitude in density [$10^{4}M_{\odot} / (10^{6} R_{\odot})^{3} \to M_{\odot}/R_{\odot}^{3}$] and roughly 11 orders of magnitude in timescale (from the dynamical time of a GMC, $\sim$1 million years, to the timescale for sound-waves in the Sun of order a few minutes). Thus the computational challenge is extreme, and one cannot hope to model star formation, using for example, uniform grid techniques, though the development of such methods are well advanced. Furthermore the physics of star formation is far from simple, involving self-gravitating, turbulent gas dynamics over a huge range of length and timescales with important contributions from magnetic fields --- ultimately requiring non-ideal magnetohydrodynamics ---, radiation transport in both optically thick and thin regimes, dust and molecular chemistry and many other physical effects which should be incorporated into a realistic model.

 Smoothed Particle Hydrodynamics \citep[SPH, for recent reviews see][]{price04,monaghan05} is a method very well suited to star formation studies because the resolution follows the mass by discretising the fluid equations onto Lagrangian particles that follow the fluid motion. The Lagrangian formulation, giving an exact treatment of mass advection, means that important conservation properties such as that of angular momentum are automatically satisfied, which requires very high resolution in grid techniques, even using adaptive mesh refinement (AMR). Techniques for solving the equations of self-gravitating hydrodynamics over large ranges of length and time in SPH are well established, based on the tree code algorithms used in $N-$body codes. By contrast, the solution of the equations of magnetohydrodynamics (MHD) in SPH has proved more challenging, in part due to the early discovery \citep{pm85} of numerical instabilities associated with particular formulations of the MHD equations. Our goal over the last few years has been to develop the techniques for MHD in SPH sufficiently to be able to study the role of magnetic fields in star formation problems. This has involved dealing carefully with many of the numerical issues including the aforementioned instabilities \citep{pm05}, the treatment of MHD shocks \citep{pm04a,pm04b} and perhaps most importantly (and the main difficulty), exploring methods for enforcing the $\nabla \cdot {\bf B} = 0$ ``no monopoles'' constraint with sufficient accuracy to perform calculations which evolve beyond the actual point of star formation \citep{pm05,pb07,price10}.

 The method we have found for enforcing the divergence constraint that is sufficiently robust for star formation studies has been to use the `Euler Potentials' or `Clebsch' formulation, whereby the magnetic field is written in terms of two scalar potentials in the form ${\bf B} = \nabla\alpha \times \nabla\beta$. The corresponding induction equation for the magnetic field takes the form
\begin{equation}
\frac{d\alpha}{dt} = 0; \hspace{1cm} \frac{d\beta}{dt} = 0,
\label{eq:ind}
\end{equation}
corresponding to the advection of magnetic field lines by Lagrangian particles. The Euler potentials are thus very naturally suited to a Lagrangian description, but there are important limitations to their use. The main one is that fields with complicated topologies (such as a poloidal field wrapped by a toroidal one) cannot be represented by Euler potentials because they would become double-valued. A corollary to this is that such fields also cannot be generated during the calculation and thus important dynamo processes are not captured. Another way to understand this is to appreciate that evolution of a field using (\ref{eq:ind}) is, in effect, a mapping of the field from the initial to final positions of the SPH particles, and requires a one-to-one mapping, after which the field winding will no longer be captured. A further issue is that it is difficult to formulate non-ideal MHD terms for the Euler potentials --- although we add artificial dissipative terms to capture shocks it is clear that these do not and cannot be used to represent a correct physical dissipation \citep[see][]{brandenburg10}.

 Nevertheless, with the above caveats in mind, we have been able to study the effect of magnetic fields on the star formation process, mainly studying the influence of the magnetic field in supporting the cloud in the initial stages of collapse, and the effect of this on the subsequent star formation sequence. Rather than starting with global turbulent-cloud star cluster formation calculations, we first studied the effect of magnetic fields on the formation of individual stars at small scales, from which we have proceeded to study star cluster formation on larger scales (see following sections).
 
 Alongside the development of the MHD algorithms, we have also developed an algorithm for incorporating the effect of radiation using the flux-limited diffusion approximation. This is an approximation in radiation is assumed to be transported by diffusion through both optically thick and thin regions, but with the diffusion speed limited to the speed-of-light in optically thin regimes. The key challenge for adapting grid-based flux-limited diffusion techniques into an SPH context was to develop an implicit integration method that enables the radiative transport to (which is much faster than the gas dynamics, particularly in optically thin regions) to be computed on a timescale similar to the hydrodynamics \cite[for details see][]{wb04,wb05}.

\begin{figure}
\centering
\includegraphics[angle=90,width=\textwidth]{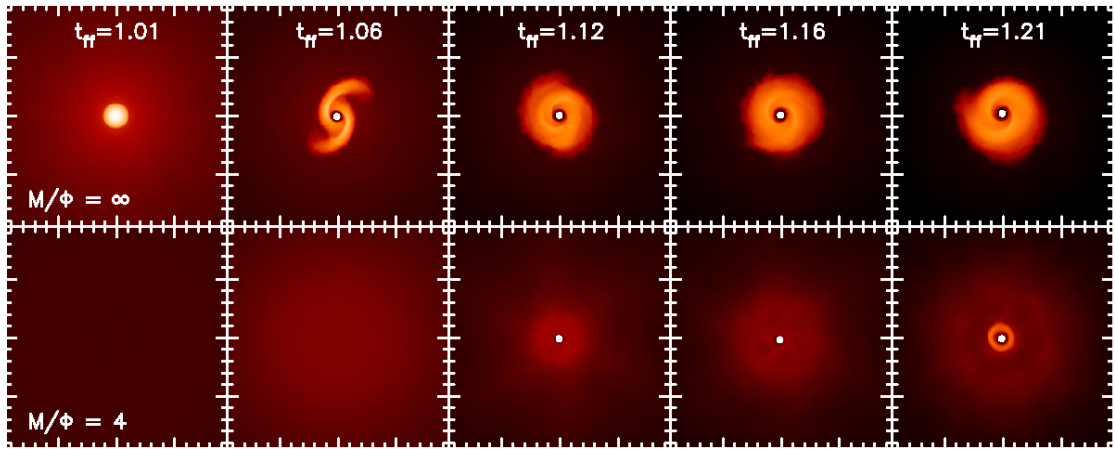}
\caption{Effect of magnetic fields on the formation of circumstellar discs around young stars: Results of a simulation following the collapse of a rotating $1M_{\odot}$ spherical cloud core without (top) and with (bottom) a uniform magnetic field threading the initial cloud. Field strengths are given in terms of the mass-to-magnetic flux ratio divided by the critical value at which magnetic fields prevent collapse altogether. Despite the relatively weak field with respect to gravity the magnetic field is able to almost completely prevent disc formation due to a combination of magnetic braking and magnetic pressure in the collapsing core. Time is shown in units of the gravitational free-fall time ($t_{ff}$).}
\label{fig:discs}
\end{figure}

\section{Single and binary star formation}
 Our first application of our MHD-SPH algorithm was to the collapse of a $1M_{\odot}$, $R=4\times 10^{16}$cm molecular cloud core to form single and binary stars. As the initial condition we assumed a dense, spherical, cold ($T\sim 10K$) core in pressure-equilibrium with a warm, low density medium with an initially uniform magnetic field threading the core and the medium. The sphere was given an initial solid body rotation, of $\Omega = 1.77 \times 10^{-13}$ rad s$^{-1}$ for the case of a single star, and $\Omega = 10^{-12}$ rad s$^{-1}$ for a binary, in the latter case also imposing an initial $m=2$ perturbation in density to seed the binary formation. We considered a range of field strengths from zero up to the observed mass-to-flux ratios of a few. The efficient cooling of the molecular gas was modelled using a barotropic (pressure-dependent-on-density) equation of state that assumes isothermality whilst the density is below a threshold physical value ($\rho_{c}=10^{-14}$ g/cm$^{3}$) and becomes polytropic with $\gamma=1.4$ above the critical density in order to approximate the effect of the gas becoming optically thick to radiation. Above a particular (much higher) density and provided the gas is gravitationally bound and collapsing, a `sink' particle is inserted to replace the densest, bound gas in the calculation, such that the calculations can be followed beyond the point of actual star formation.
 
  The results of a typical set of calculations are shown in Fig.~\ref{fig:discs}, showing the projected column density as it evolves beyond one gravitational free-fall time (left-to-right, times given in units of $t/t_{ff}$), for a calculation with no magnetic fields (top) and with a mass-to-flux ratio of 4 times the critical value --- that is magnetic fields that are too weak to prevent collapse by a factor of 4). Despite the relatively weak field the effect on the formation of the circumstellar disc is catastrophic. In the absence of a magnetic field a disc is formed that is sufficiently massive so as to become unstable to gravitational instability in the form of large scale spiral arms, yet with a magnetic field only the faintest trace of a disc is visible even at the end of the calculations.
  
   Since our initial calculations by several other groups have found similar results based on numerical simulations \citep[e.g.][]{ht08,hc09} and also semi-analytic calculations by \citet{gallietal06} (see Galli, this volume). In fact \cite{ht08} somewhat alarmingly discuss a `fragmentation crisis' and speculate further that, given the paucity of observational evidence for discs in the earliest (class 0) phase of star formation, perhaps they do not exist (instead forming later). More likely the solution lies in the fact that we have assumed ideal MHD in a regime where it is clear that non-ideal MHD effects are known to be important. Indeed later analysis by \citet{shuetal06} suggests that Ohmic resistivity can provide a solution, though nonetheless requiring a diffusion parameter considerably higher than the microscopic value. We intend to explore non-ideal MHD effects in the near future, though it requires a shift away from the Euler potentials formulation \cite[for recent progress on this, see][]{price10}.

 A similarly dramatic effect of magnetic fields on binary formation was also observed, though for the case of binaries the effect depended more strongly on the magnetic field configuration, since in certain circumstances the field configuration could assist binary formation by forming a ``magnetic cushion'' between two overdense, collapsing regions. It was also found that a sufficiently large perturbation would produce a binary regardless of the magnetic field strength. Nevertheless it is clear that the presence of even a relatively weak magnetic field in a molecular cloud core can drastically change the star formation picture.

\section{Effect of magnetic fields on cluster formation}
 We have also considered the effect of magnetic fields on larger scales, important to the formation of whole star clusters \citep{pb08,pb09}. Our initial study was to evaluate the influence of magnetic fields in star cluster formation calculations similar to those performed by \citet{bbb03}. The initial conditions consist of a cold ($T=10K$), 50$M_{\odot}$, uniform density cloud of radius $\sim 0.2$ pc, with an imposed turbulent velocity field with a power spectrum and Mach number $\mathcal{M} = 6.7$ consistent with observed motions in molecular clouds. As previously we initially adopted a barotropic equation of state (see above) to approximate the effect of the gas becoming optically thick, and thus heating and halting the collapse, beyond a certain critical density. The initially turbulent cloud was threaded with a uniform magnetic field, though with no external medium. Instead, SPH particles in the initially expanding outer layers of the cloud carry the magnetic field outwards and form a low density medium into which the field is anchored.
 
\begin{figure}
\centering
\includegraphics[width=\textwidth]{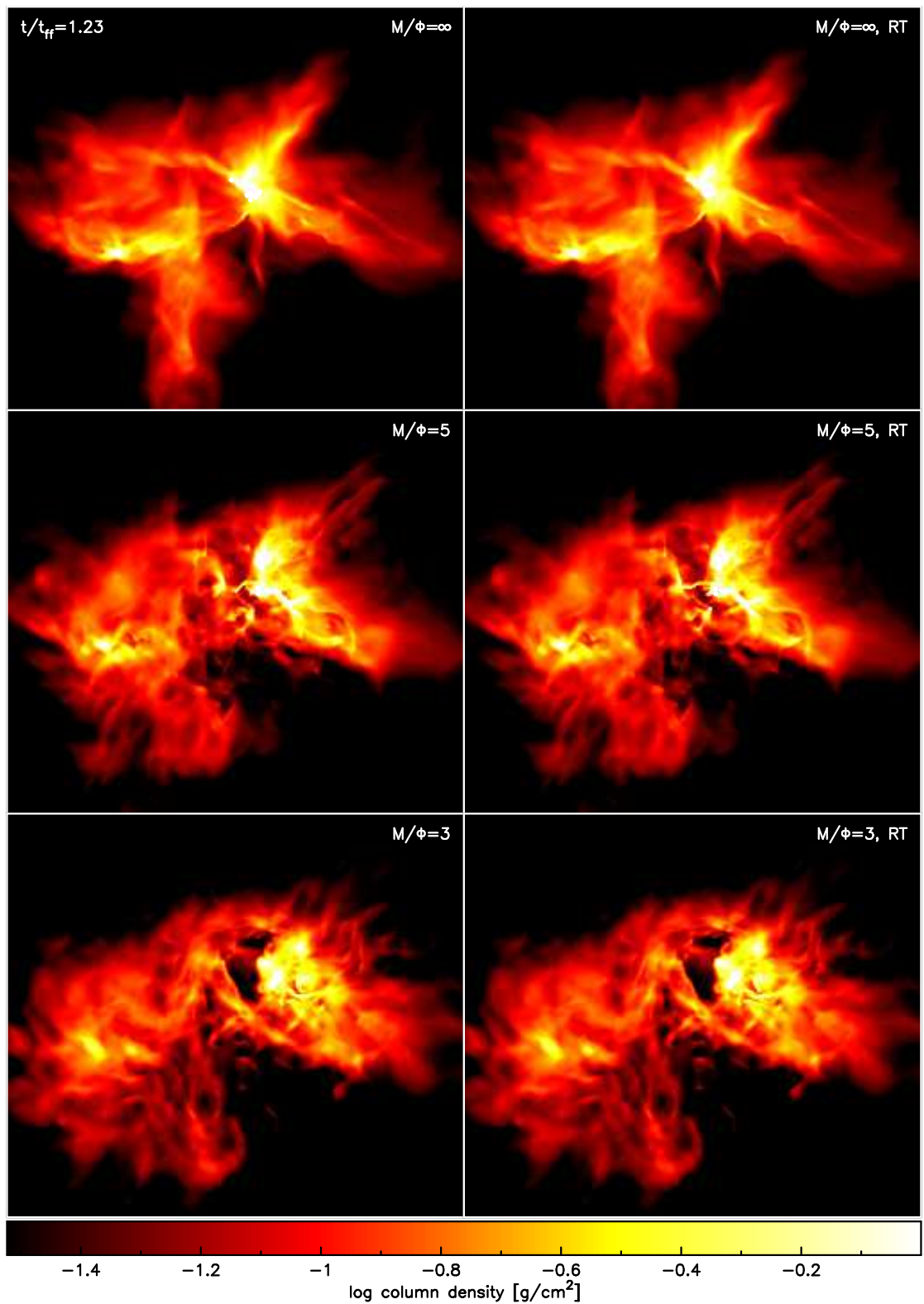}
\caption{Effect of magnetic fields and radiation on the large scale structure of star-forming 50 $M_{\odot}$ molecular cloud cores. Showing calculations with no magnetic fields (top row), a mass-to-flux ratio of 5 (middle row) and 3 times the critical value (bottom row). In the regime where magnetic pressure exceeds gas pressure the magnetic fields there is a dramatic influence on the global cloud structure, with the appearance of large-scale, magnetically supported voids. The large scale evolution of the cloud with radiative transfer explicitly calculated (right panels) is identical to that using an approximate, barotropic equation of state (left panels), at least for low mass star formation. }
\label{fig:global}
\end{figure}

 Despite the relatively weak field strengths with respect to gravity, magnetic fields were found to have a dramatic effect on the large scale structure of the clouds, as can be seen from the column density projection shown at $t/t_{ff} = 1.23$ in Fig.~\ref{fig:global}. This is because the fields are \emph{not} weak with respect to gas pressure, so the magnetic field is able to act as the dominant source of pressure within the cloud, producing large-scale magnetically-supported voids (middle and bottom rows) that are completely absent from purely hydrodynamical calculations (top row).
 
 The means by which magnetic fields are able to act as a source of pressure on large scales is relatively simple to understand, since in ideal MHD the gas motions are tied to the magnetic field lines. For a relatively strong field, this means that gas is channelled along field lines as it collapses (rather than the gas dragging the field lines around in the weak field case). Since in ideal MHD the mass-to-flux ratio is conserved along any given flux tube, any gas collapsing to form dense structures inevitably leaves behind a region evacuated of gas pressure but with the magnetic field strength (and magnetic pressure) unchanged. Thus the ratio of gas-to-magnetic pressure decreases substantially away from the densest gas. New material is prevented from re-entering the evacuated region because of the inability to cross magnetic field lines. Thus the region, once evacuated, remains as a magnetic-pressure supported void. At a recent meeting the above mechanism was paraphrased by Carl Heiles as ``magnetic fields abhor a vacuum'', since it is easy to remove gas from a region of space along the magnetic field lines, but the magnetic fields themselves will remain.

\begin{figure}
\centering
\includegraphics[width=0.8\textwidth]{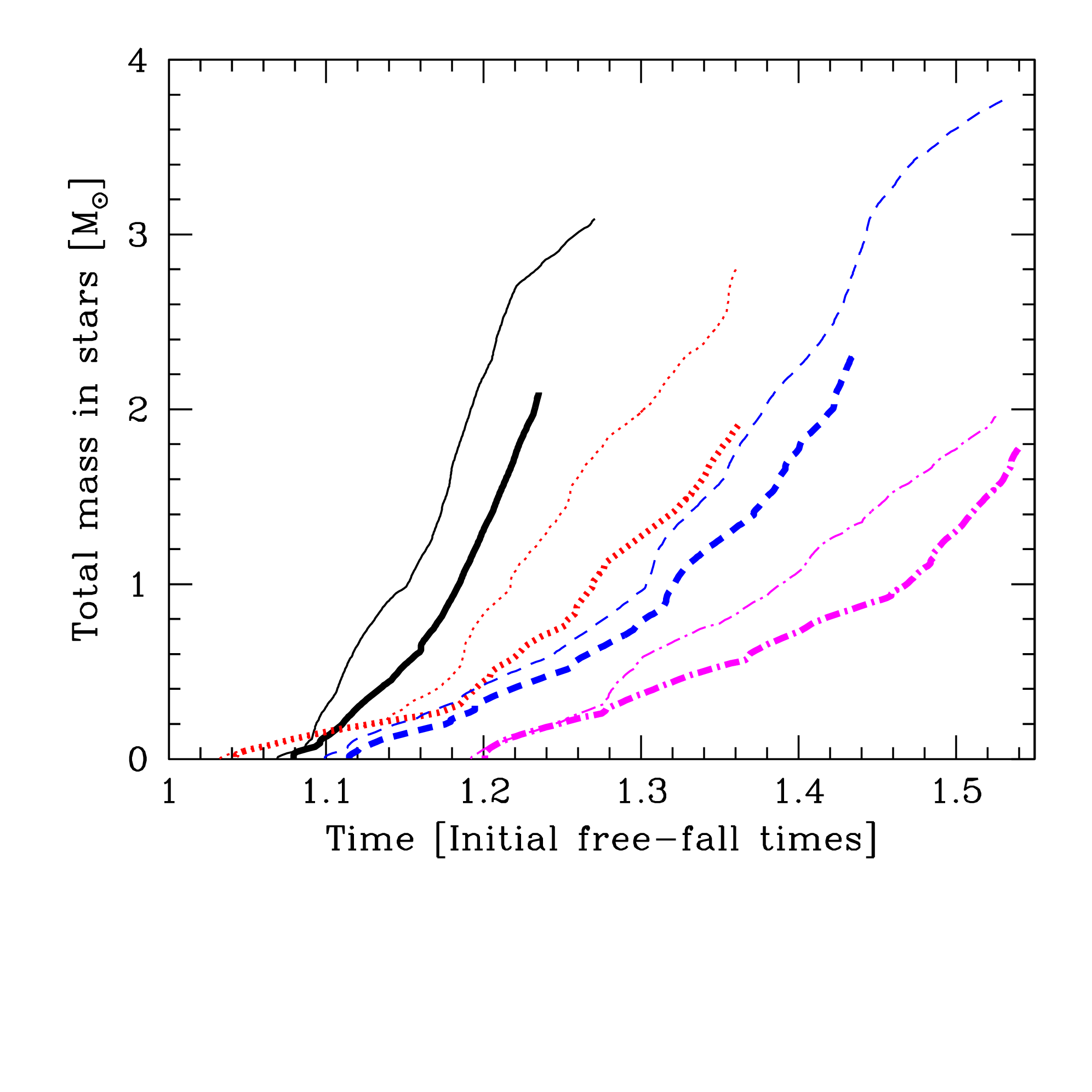}
\caption{Combined influence of magnetic fields and radiative feedback on the star formation rate in our $50 M_{\odot}$ model clouds. Line styles correspond to the four different magnetic field strengths employed: no magnetic fields (solid, black), and mass-to-flux ratios of 10, 5 and 3 in units of the critical value for preventing collapse altogether (dotted red, dashed blue and dot-dashed magenta lines respectively), whilst the line width shows whether (thick lines) or not (thin lines) radiative feedback was modelled (if not, a barotropic equation of state was employed). The magnetic field strength has the dominant influence on the star formation rate, with a secondary effect due to radiative feedback occurring at later times.}
\label{fig:sfr}
\end{figure}

 The effect of the support provided to the large scale regions of the cloud is a dramatic slow-down in the star formation rate with increasing magnetic field strength (discussed below, see Fig.~\ref{fig:sfr}), most effective in the regime where magnetic pressure exceeds gas pressure ($\beta < 1$) and independent of the fact that the field may be weak relative to gravity and/or turbulence. An unexpected finding from \citep{pb08} was the resultant change to the initial mass function of stars formed in the calculations, in the form of a reduction in the number of sub-stellar objects (i.e., brown dwarfs) relative to higher mass objects (i.e., stars). This occurs not because of some complicated influence of the magnetic fields on the fragmentation --- we do not resolve the magnetic fields structure on the smallest scales in these calculations --- but simply because of the overall slowdown in the star formation rate and a consequent reduction in the importance of dynamical interactions and the associated ejection of low mass objects from multiple systems. Given the low number of objects formed overall in the strong magnetic field calculations, it is not possible to state whether or not this effect is sufficient to resolve the statistical disagreement in the number of low mass objects and the observed IMF found by \citet{bate09a}, but the trend is certainly in the right direction.
 
 In the strongest field calculation, we also found that the expanding outer regions of the cloud started to show a `striped' appearance as the gas was channelled along the magnetic field lines. This is strongly reminiscent of the `magnetically aligned striations' observed in the outer regions of the Taurus molecular cloud in the recent $^{12}$CO and $^{13}$CO molecular line emission maps by \citet{goldsmithetal08}, co-located with measurable velocity anisotropy aligned with the global magnetic field \citep{heyeretal08}. This is a good indication that Taurus lies in a regime where the magnetic field is able to exert considerable influence on the star formation process.

\begin{figure}
\centering
\includegraphics[width=\textwidth]{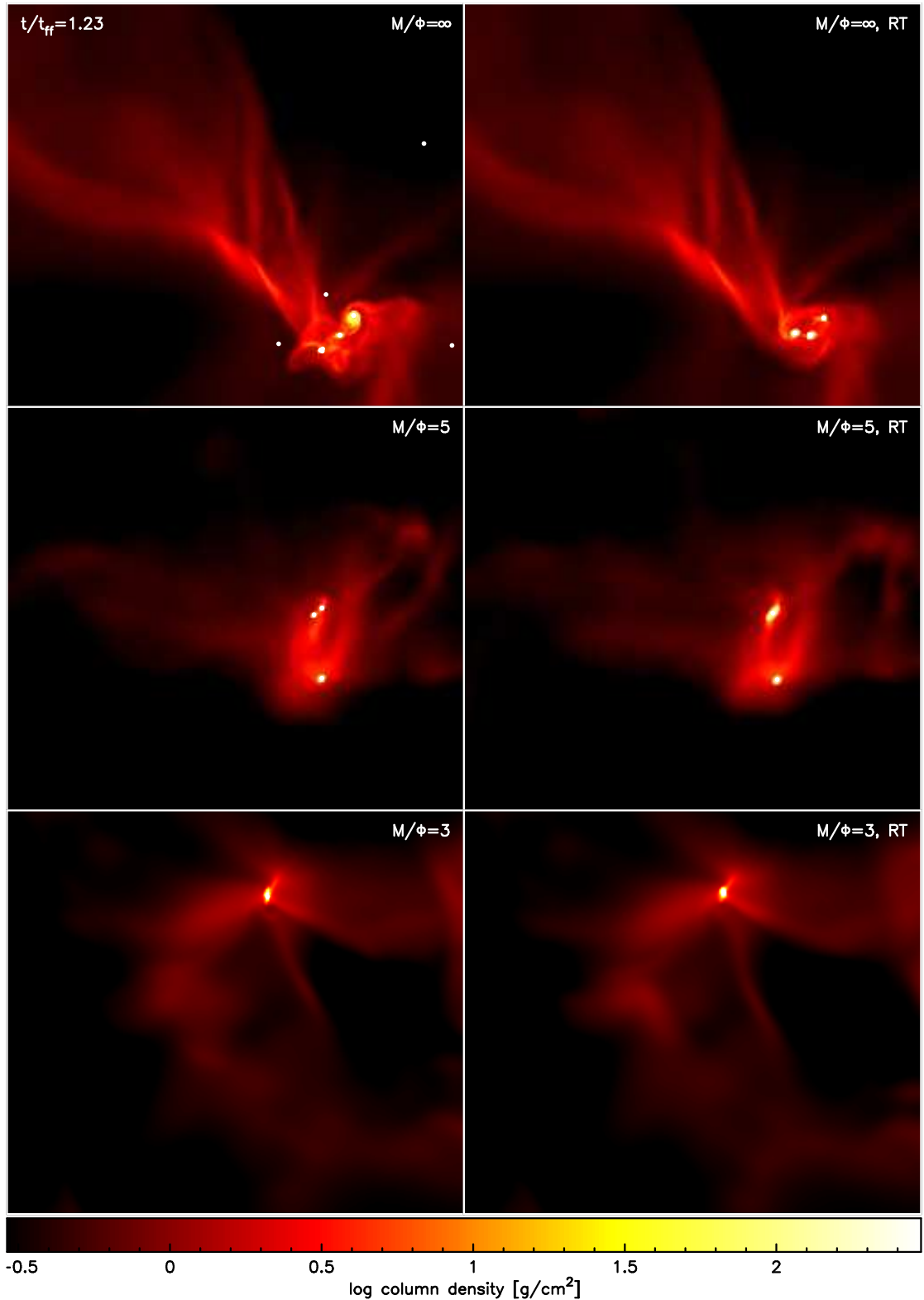}
\caption{Combined effect of magnetic fields and radiative feedback on star cluster formation. The plots show a zoomed-in subsection of the clouds shown in Fig.~\ref{fig:global} at 1.23 initial gravitational free-fall times for the calculations of three different magnetic field strengths (top to bottom), without (left) and with (right) a full modelling of radiative transport in the gas. The effect on the small scale fragmentation is dramatic: Once the gas becomes optically thick to radiation the heating effect provided to neighbouring material completely inhibits any subsequent fragmentation within a radius of several AU. The result is a trend towards fewer but more massive stars and a further reduction in the overall star formation rate on top of the large-scale effect provided by the magnetic field.}
\label{fig:zoomin}
\end{figure}

\section{Influence of radiative feedback on star cluster formation}
 A key limitation to all of the calculations discussed above was the approximate treatment of the thermodynamics of the gas via the use of a barotropic equation of state where gas pressure is a function of density alone, rather than being a function of density and temperature. Naturally this assumption simplifies the calculations considerably, but it misses important feedback processes, especially once the gas enters the optically thick regime. In particular, using the barotropic approximation the temperature is assumed to rise strictly with density, but this neglects the fact that radiation in actual fact should diffuse from the hot, dense, compressed gas into the less-dense surrounds, thus heating it and preventing it from fragmenting further.
 
  In the initial phases of the cloud evolution and on the largest scales the radiation has very little influence, evident from Fig.~\ref{fig:global} which compares the cloud structures using the barotropic equation of state (left panels) with calculations incorporating the transport of radiation within the gas via the flux-limited diffusion approximation (right panels). This is partly the case because we form only low-mass stars in the calculations, but also because we have neglected both the accretion luminosity within the sink radius and the luminosity of the protostars themselves. Thus the effect of radiation that we consider is very much a lower limit to the true feedback effect. However, to capture as much of the radiative feedback as possible we have reduced the accretion radius on the sink particles to a mere $0.5$ AU in size, compared to $5$ AU sink radii in previous calculations \citep[e.g.][]{bbb03,pb08}. Other authors \citep{offneretal09} adopt a prescription for providing radiative feedback from the protostars themselves, but this brings a number of assumptions about the protostellar evolution process that at this stage we have preferred to avoid since it ultimately requires a detailed sub-grid stellar evolution model in order to constrain the parameters.
  
   The effect provided by the transport of radiation from the hot, collapsing, compressed gas into neighbouring material does however have a dramatic impact on the small-scale fragmentation in the cloud. This is illustrated in Fig.~\ref{fig:zoomin} that shows a close up view of the star formation occurring in the 6 model clouds shown in Fig.~\ref{fig:global}. Whilst the effect of the magnetic field on the large scales is to reduce the overall rate at which the global cloud collapses, radiative feedback completely inhibits any secondary fragmentation in the material immediately surrounding the protostars (comparing left to right panels). The result is a reduction in the formation of low mass objects (visibly being ejected from the multiple systems resulting in the top left panel), particularly those that initially resulted from fragmentation in circumstellar discs which with the radiative feedback effect included become sufficiently heated such that no further sub-fragmentation occurs. Similarly in the middle panels at moderate field strength it may be observed that an object that initially fragmented into a binary system using the barotropic equation of state no longer fragments when radiative transport is accounted for, instead producing a single star with a circumstellar disc.
 
\begin{figure}
\centering
\includegraphics[angle=270,width=\textwidth]{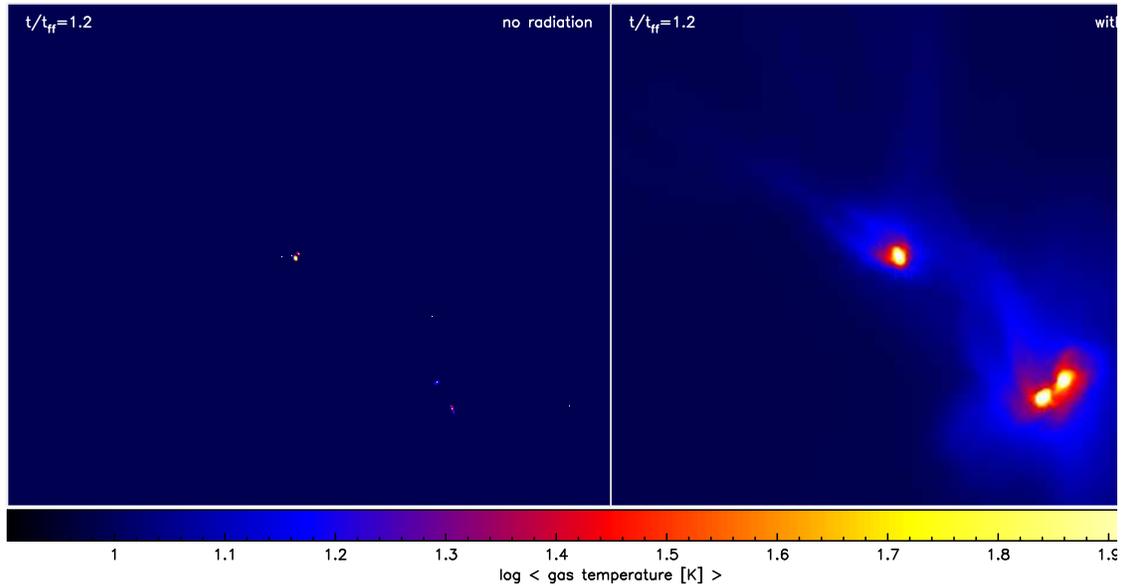}
\caption{Effect of radiative feedback on star cluster formation: The plot shows the distribution of average temperature $\int \rho T {\rm dz}/ \int \rho {\rm dz}$ from the hydrodynamic calculations shown in the top row of Figs.~\ref{fig:global} and \ref{fig:zoomin} employing a barotropic (pressure-proportional to density) equation of state (left panel) compared to the calculation (right) where the radiation is explicitly modelled and thus the transport of radiation from hot to cold regions is captured. Whereas using the barotropic approximation only the material above the critical density becomes hot, in the radiation hydrodynamics calculation an extended region surrounding each protostar is heated and thus fragments no further (see Fig.~\ref{fig:zoomin}).}
\label{fig:temp}
\end{figure}

 The reason for the dramatic reduction in small scale fragmentation that occurs when radiative feedback is easily understood from our above discussion, and may be readily illustrated by a plot of the integrated temperature $\int \rho T {\rm dz}/ \int \rho {\rm dz}$ shown in Fig.~\ref{fig:temp} for the hydrodynamic calculations using the barotropic equation of state (left panel) and with radiative feedback included (right panel). With the barotropic equation of state (left) the temperature is high only at several discrete points corresponding to where the gas density exceeds the threshold for the polytropic index to change from $\gamma = 1$ to $\gamma = 1.4$. Once the transport of radiation from hot to cold regions is modelled (right panel), a spatially extended region of high temperature gas is produced in the region surrounding each collapsing protostar, producing the effect on the fragmentation seen in Fig.~\ref{fig:zoomin}.
Our results regarding the effect on small-scale fragmentation produced by the transport of radiation have been confirmed by calculations employing similar physics performed by \citet{offneretal09} using an adaptive mesh refinement code.

 The resultant effect on the initial mass function strengthens the trend already produced by the magnetic field, namely towards producing fewer and more massive objects. As previously stated, this is in the right direction to resolve the discrepancy with the observed IMF found by \citet{bate09a} in barotropic calculations, but given the very low number statistics --- particularly with the overall reduction in star formation rate produced by the combined influence of the magnetic fields and radiation --- we are reluctant to draw firm conclusions in this regard.

\section{Combined influence of magnetic field and radiative feedback on the star formation rate and efficiency}
 Having assessed the effect of both magnetic fields and radiative feedback on the star cluster formation, we may return to our original question: namely, are these two pieces of missing physics the necessary and sufficient ingredients required to explain the kind of slow and inefficient star formation observed in real molecular clouds?
 
  The combined effect on the star formation rate is shown in Fig.~\ref{fig:sfr}, showing the total mass accreted onto the sink particles for the full suite of calculations as a function of time. The primary influence on the star formation rate is the strength of the initial magnetic field, since it affects the large scale structure of the cloud and thus the amount of material that is able to later collapse and form stars. Radiative feedback enters as a secondary effect, reducing the star formation rate further, particularly at later times as the radiation diffuses further from the protostars into the surrounding medium.
  
  It is notable that only the calculations employing the strongest magnetic fields (mass-to-flux ratio of 3 in units of the critical value) produce a star formation rate that is even remotely close to the observed rate of 3-6\% per gravitational free-fall time found by \citet{evansetal09}: The rate in the strongest field calculation with radiative feedback is $0.18 M_{\odot} / 0.34 t_{ff} /50 M_{\odot} \approx 10\%$ per free-fall time. This is not unreasonable since molecular cloud cores are indeed observed to have mass-to-flux ratios of a few times the critical value, and radiative feedback is clearly an important effect. Relative differences in the star formation rate across the Galaxy can also be explained as being due to variations in the global flux threading individual star forming clouds.
We can speculate that the remaining discrepancy between our results and observational estimates of the star formation rate is due to our neglect of additional feedback processes, namely the intrinsic and accretion luminosity from the protostars themselves as well as mechanical feedback from jets and outflows.
  
  The question of the overall star formation \emph{efficiency} in the presence of magnetic fields and radiation transport is more difficult to answer given the limited time for which we have been able to evolve the calculations beyond one free-fall time. Ideally one would continue the calculations over several global dynamical times until star formation activity has ceased, however this is currently prohibitively expensive in terms of CPU time. Observational estimates are limited in a similar manner because a star forming molecular cloud is defined as one in which star formation has initiated but not completed, and once completed one has little insight as to the initial mass of the parental cloud.    If we assume that star formation continues at the rate observed in Fig.~\ref{fig:sfr} and that the molecular cloud survives for 2-3 free-fall times beyond star formation, then the overall star formation efficiency in the strongest field case would be of order 20-30\%. By contrast, for the calculations without magnetic fields the efficiency would be close to 100\% on a similar timescale. Since at supercritical mass-to-flux ratios the field is relatively weak compared to gravity, the fraction of bound gas at the end of the calculation remains relatively high even for the highest field strengths, of order 85\% for the mass-to-flux ratio of 3 (times critical) calculation with radiative feedback, so the main requirement for a low overall efficiency is that the cloud should be dispersed after several dynamical times and that the star formation rate should not accelerate considerably with time.

\section*{Acknowledgements}
DJP is supported by a Monash Fellowship, though much of this work was completed whilst funded by a UK Royal Society University Research Fellowship at the University of Exeter. We thank the organisers for their hospitality in both Milan and Como, the opportunity to attend and present at the conference.

\bibliography{sph,starformation,mhd}

\end{document}